\documentclass[prl,twocolumn,showpacs,superscriptaddress]{revtex4}
\usepackage{amsmath,amssymb,graphicx,epstopdf,color,bm,soul,upgreek, mathtools}

\begin{document}

\title{Photon number-resolved measurement of an exciton-polariton condensate}

\author{M. Klaas}
\thanks{These two authors contributed equally}
\affiliation{Technische Physik, Wilhelm-Conrad-R\"ontgen-Research Center for Complex
Material Systems, Universit\"at W\"urzburg, Am Hubland, D-97074 W\"urzburg,
Germany}
\author{E. Schlottmann}
\thanks{These two authors contributed equally}
\affiliation{Institut f\"ur Festk\"orperphysik, Quantum Devices Group, Technische Universit\"at Berlin,
Hardenbergstra\ss e 36, D-10623, 10623 Berlin, Germany}
\author{H. Flayac}
\affiliation{Institute of Physics, Ecole Polytechnique Federale de Lausanne (EPFL), CH-1015 Lausanne, Switzerland}
\author{F. P. Laussy}
\affiliation{Faculty of Science and Engineering, University of Wolverhampton, Wulfruna St, Wolverhampton WV1 1LY, UK}
\affiliation{Russian Quantum Center, Novaya 100, 143025 Skolkovo, Moscow Region, Russia}
\author{F. Gericke}
\affiliation{Institut f\"ur Festk\"orperphysik, Quantum Devices Group, Technische Universit\"at Berlin,
Hardenbergstra\ss e 36, D-10623, 10623 Berlin, Germany}
\author{M. Schmidt}
\affiliation{Institut f\"ur Festk\"orperphysik, Quantum Devices Group, Technische Universit\"at Berlin,
Hardenbergstra\ss e 36, D-10623, 10623 Berlin, Germany}
\affiliation{Physikalisch-Technische Bundesanstalt,  Abbestrasse 2-12, 10587 Berlin, Germany}
\author{M. v. Helversen}
\affiliation{Institut f\"ur Festk\"orperphysik, Quantum Devices Group, Technische Universit\"at Berlin,
Hardenbergstra\ss e 36, D-10623, 10623 Berlin, Germany}
\author{J. Beyer}
\affiliation{Physikalisch-Technische Bundesanstalt,  Abbestrasse 2-12, 10587 Berlin, Germany}
\author{S. Brodbeck}
\affiliation{Technische Physik, Wilhelm-Conrad-R\"ontgen-Research Center for Complex
Material Systems, Universit\"at W\"urzburg, Am Hubland, D-97074 W\"urzburg, Germany}
\author{H. Suchomel}
\affiliation{Technische Physik, Wilhelm-Conrad-R\"ontgen-Research Center for Complex
Material Systems, Universit\"at W\"urzburg, Am Hubland, D-97074 W\"urzburg, Germany}
\author{S. H\"ofling}
\affiliation{Technische Physik, Wilhelm-Conrad-R\"ontgen-Research Center for Complex
Material Systems, Universit\"at W\"urzburg, Am Hubland, D-97074 W\"urzburg, Germany}
\affiliation{SUPA, School of Physics and Astronomy, University of St Andrews, St Andrews
KY16 9SS, United Kingdom}
\author{S. Reitzenstein}
\affiliation{Institut f\"ur Festk\"orperphysik, Quantum Devices Group, Technische Universit\"at Berlin,
Hardenbergstra\ss e 36, D-10623, 10623 Berlin, Germany}
\author{C. Schneider}
\affiliation{Technische Physik, Wilhelm-Conrad-R\"ontgen-Research Center for Complex
Material Systems, Universit\"at W\"urzburg, Am Hubland, D-97074 W\"urzburg, Germany}

\begin{abstract}
We measure the full photon-number distribution emitted from a Bose condensate of microcavity exciton-polaritons confined in a micropillar cavity. The statistics are acquired by means of a photon-number resolving transition edge sensor. We directly observe that the photon-number distribution evolves with the non-resonant optical excitation power from geometric to quasi-Poissonian statistics, which is canonical for a transition from a thermal to a coherent state. Moreover, the photon-number distribution allows evaluating the higher-order photon correlations, shedding further light on the coherence formation and phase transition of the polariton condensate. The experimental data is analyzed in terms of thermal coherent states which allows one to directly extract the thermal and coherent fraction from the measured distributions. These results pave the way for a full understanding of the contribution of interactions in light-matter condensates in the coherence buildup at threshold.
\end{abstract}

\maketitle

Quantum condensation, in the case of photonic systems \cite{Legget.2001}, describes the transition from a chaotic or thermal state of many particles to a coherent state that provides the order parameter for a macroscopic wavefunction.  This is best described by the full particle-number distribution, that embeds the correlation at all orders, while experiments usually focus on the first and second order.  The textbook case reduces to an exponential distribution for the particle number in the incoherent phase, as opposed to a Poissonian distribution in the condensed phase.  Every system, however, makes this transition in a way that is specific to
its mechanism of coherence buildup and to the conditions in which this happens. In lasers, which have been the first and foremost systems to grow coherence, quantum theory describes this transition with nonlinear master equations that include positive feedback and pumping. A popular model, the Scully-Lamb master equation, finds a
transition from a thermal state below threshold, to a bell-shaped
photon distribution above threshold, but with a much higher spread
than a Poisson distribution~\cite{scully67a}.  The measurement of the
full photon-number distribution was performed shortly after the realisation of
lasers to confirm the nature of the light field through an excellent
agreement with the ideal distributions when far enough from the
threshold~\cite{arecchi65a}. In atomic condensates, the need for a full particle-number distribution is even more compelling, as a strongly-correlated gas has a richer physics of higher-order correlations~\cite{bloch08a} that impacts on such critical dynamics as the rate of many-body collision~\cite{burt97a} or nonlocal interactions and entanglement in sufficiently interacting systems (such as those of reduced dimensionality)~\cite{gritsev06a}. The full atom-number distribution of a Bose-Einstein condensate (BEC) was also measured shortly after the system was realized in the laboratory~\cite{ottl05a}, in the atom laser configuration where the condensate is left to free-fall. The deviations from thermal and Poissonian distributions on both sides of the transitions have been found to be more important than for the photonic case, due to atomic interactions.  Full
particle-distributions have also been reported in other quantum
systems, such as superconducting qubits~\cite{schuster07a}, where the system is so-strongly quantized that its photon-statistics manifests
directly in the signal. Recently, particle-number
distributions have been used to investigate more intricate aspects of
quantum thermodynamics, such as revealing the so-called
prethermalization stage in out-of-equilibrium systems~\cite{adusmith13a} or in characterizing condensation in a closed or open system thermalizing with or without fluctuations of particles in its reservoir~\cite{schmitt14a}. In all these cases, the particle-number distributions are a precious tool to provide a comprehensive picture of the quantum state of the system, well beyond the standard correlation functions. While photonic systems on the one hand, and material ones on the other hand have been readily characterized in this way, such a characterization has been missing for another platform which also thrives with condensation phenomena, namely, exciton-polaritons. These particles are a mixture of light and matter and whether their condensation follows more the light paradigm or the atomic paradigm has been a topic of intense debates, which are still largely unresolved to this day \cite{Byrnes.2014}.
 
However, an accurate measurement of the photon probability distribution, in particular in nanoscopic-lightsources with comparatively low photon-numbers per emitted pulse, is non-trivial, as it requires in principle single-photon detection capabilities in combination with photon-number resolution. The advent of transition edge sensors (TES), which are highly sensitive calorimetric sensors in the single-photon regime, allows overcoming past difficulties \cite{Cabrera.1998}.
Their functionality relies on a temperature change at the superconducting-to-normal-conducting transition and consequent resistivity change due to the absorption of a countable number of photons \cite{Irwin.2005, Heindel.2017}. Modern transition edge sensors can moreover exhibit a near-unity detection efficiency over a wide range of wavelengths, which makes them a highly versatile tool for the characterization of light-sources \cite{Lita.2008}, including micro- and nano-laser devices operated in the few photon regime. 

Strong coupling conditions leading to the formation of exciton-polaritons in quantum well-microcavities, have first been observed by Weisbuch \textit{et al.} \cite{Weisbuch.1992}. In the high-density regime, the system can undergo a transition to a dynamic Bose-Einstein condensate aided by bosonic final state stimulation \cite{Kasprzak.2006, Imamoglu.1996}. As a result of the driven-dissipative nature of the system, the coherence properties of such condensates can be investigated by studying the properties of the (spontaneously) emitted photons in the spatial and temporal domain. A variety of studies, relying on Michelson interferometry and double-slit experiments, have focussed on adressing spatial coherence properties through ($g^{(1)}$(r)) measurments \cite{Deng.2007, Roumpos.2012, Fischer.2014}. Temporal coherence has been extensively studied by determining the second order autocorrelation function $g^{(2)}(\tau)$ with avalanche photodiodes \cite{Deng.2003, Kasprzak.2008, Love.2008}. The extension to three detectors has allowed to access the third-order correlation function $g^{(3)}(\tau)$~\cite{Horikiri.2010}. Finally, a special streak camera method has also been employed to resolve up to the fourth order in a semiconductor microcavity system, although not in the polariton condensation phase \cite{Assmann.2009, Assmann.2010}.

These partial measurements of the statistical properties of the emission resulted in contradictory results \cite{Kavokin.2017}. From the beginning of this research field of investigating coherence buildup mechanisms in light-matter condensates, the question regarding the importance of interactions in the phase transition arose.  Namely, it was challenged whether the condensation could result from relaxation towards the ground state mainly due to Bose stimulation of
the scattering, akin to a polariton laser~\cite{laussy04c}, or on the
opposite following instead the atomic situation ruled by interactions,
and the strong correlations that result, resuling in more marked
deviations of the polariton-number statistics~\cite{schwendimann08a}. Such questions can be answered by confronting available theories to the experiment.

In this Letter, we investigate a strongly coupled microcavity in the regime of polariton condensation via a transition edge sensor. This photon-number resolving experiment allows us to reconstruct the photon-number probability distribution $P_n$, and, consequently to assess high orders of the autocorrelation function via~\cite{Barnett.2002} $g^{(k)}(0)=\frac{\sum_n \prod_{i=0}^{k-1}(n-i)P_n}{(\sum_n n P_n )^k}$, where $g^{(k)}(0)$ denotes the autocorrelation function of $k$-th order at zero time delay, $n$ is the photon-number and $P_n$ the probability to find $n$ photons. We demonstrate that such a quantum fluid of light, which is generated in a cylindrical micropillar cavity under optical pumping, exhibits a transition that follows closely the non-interacting scenario that transits from an ideal thermal distribution to an ideal Poisson distribution. This happens without the significant departures that are observed when strong interactions play a chief role in the condensate nucleation.  Interestingly, however, and as should be expected, we still observe slight deviations that cannot be attributed to experimental error. These exist even for the ideal gas when including particle-number correlations, which is required to grow coherence~\cite{laussy12a} (rate equations alone imply thermal
statistics for all states regardless of their occupancy), and such
deviations are more pronounced nearby the threshold, suggesting that
the ones we observe in the experiment originate from the underlying
coherence buildup mechanism, thus providing precious data for further
theoretical and experimental analysis.

\begin{figure}[t]
\centering
\includegraphics[width=\linewidth]{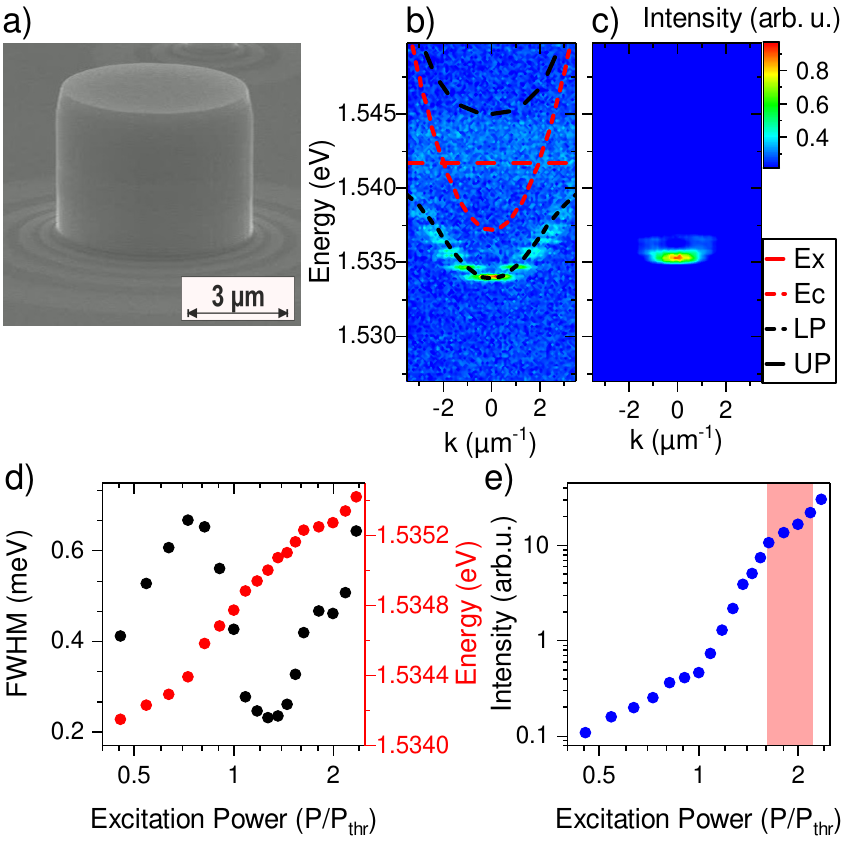}
\caption{a) Scanning electron microscopy (SEM) image of the micropillar device. b) Low power dispersion at a pump power of 0.2$P_{th}$. Red dashed parabolic/straight line signify the photon (Ec)/exciton (Ex) mode of the system. The dashed black lines is a coupled oscillator fit to calculate the upper (UP) and lower (LP) polariton branches. c) System driven above the nonlinearity into the polariton condensate regime at 2$P_{th}$. d)/e) Characteristics of energy, linewidth and intensity of the polariton emission relative to input power extracted from a Lorentzian fit of the integrated groundstate emission. The red area marks the power range investigated with the TES.}
\label{Fig1}
\end{figure}

Figure \ref{Fig1}b) depicts a low-power momentum-resolved photoluminescence measurement \cite{Lai.2007} of the micropillar at 0.2$P_{th}$, which yields the dispersion of the emission from the lower polariton branch ($k=0\,$ $\upmu$m$^{-1}$ at $1.534\,$eV). $P_{th}$ has been set to the onset of the intensity nonlinearity of the device. Due to the lateral photonic confinement in the micropillar, the modes are quantized, and we observe a mode separation of $0.630\,$meV between ground state and the first excited optical mode in the pillar cavity. A simple analytical calculation of the photonic mode splitting in a cylindrical microcavity \cite{Gutbrod.1998}, scaled with the photonic Hopfield coefficient, would result in a pillar diameter of $\approx$ 6.2 $\upmu$m for the measured mode separation. At energies of $1.5417\,$eV and $1.5438\,$eV, we observe a dispersion-less emission signal approximately at the free-exciton resonance, which is a commonly observed consequence of the photonic confinement generated via etching through the active medium \cite{Schneider.2017}. Due to the high Q-factor in our device, we do not observe any emission signal from the upper polartion branch \cite{Kulakovskii.2010}. We fit the discretized dispersion of the lower polariton branch with a standard coupled oscillator theory \cite{Kavokin.2017, Savona.1995} (Fig. 1b), which results in a detuning of $-4.5\,$meV between photon and exciton modes (red dashed lines) in our device, corresponding to a fraction of 30 \% exciton at k=$0\,$~$\upmu$m$^{-1}$. Figure \ref{Fig1}c) shows an angle-resolved measurement for elevated pump powers, approximately a factor of two above the nonlinear threshold of our device. In this regime, the emission is dominated by a monochromatic signal, which is slightly blue-shifted from the polariton ground state at low densities. In Figs. \ref{Fig1}d) and e) we investigate the emission characteristics with increased excitation power. The data has been extracted from a Lorentzian fit of the integrated ground state emission. At threshold, final state parametric scattering \cite{Tartakovskii.2002} originating in the bosonic nature of the quasi-particles begins to dictate the relaxation dynamics and we observe a macroscopically populated groundstate. This results in an intensity nonlinearity [see Fig. \ref{Fig1}d)], evidenced in the changes of the slope (s-shaped) around threshold. Furthermore, the emission energy of the mode blue-shifts with excitation power due to the excitonic fraction of the micropillar in strong coupling conditions \cite{Ciuti.1998}. The logarithmic form above the polariton phase transition has been previously reported, e.g. in Ref. \cite{Roumpos.2010} and is attributed to exciton-exciton interaction screening with increased polaritonic density. The linewidth drops sharply, which is a common sign of coherence buildup, related to the coherence time increase of the condensate via the Wiener-Khinchin theorem \cite{Wiener.1964}. After threshold, the linewidth slowly increases due to decoherence inducing particle fluctuations and interactions with the crystal environment again stemming from the part matter nature of the condensate \cite{Krizhanovskii.2006, Love.2008}. These three characteristic behaviors are commonly used to evidence persisting strong coupling conditions in a microcavity system emerging with increased particle density across its phase transition \cite{Balili.2007}. 

\begin{figure}[tbp]
\centering
\includegraphics[width=\linewidth]{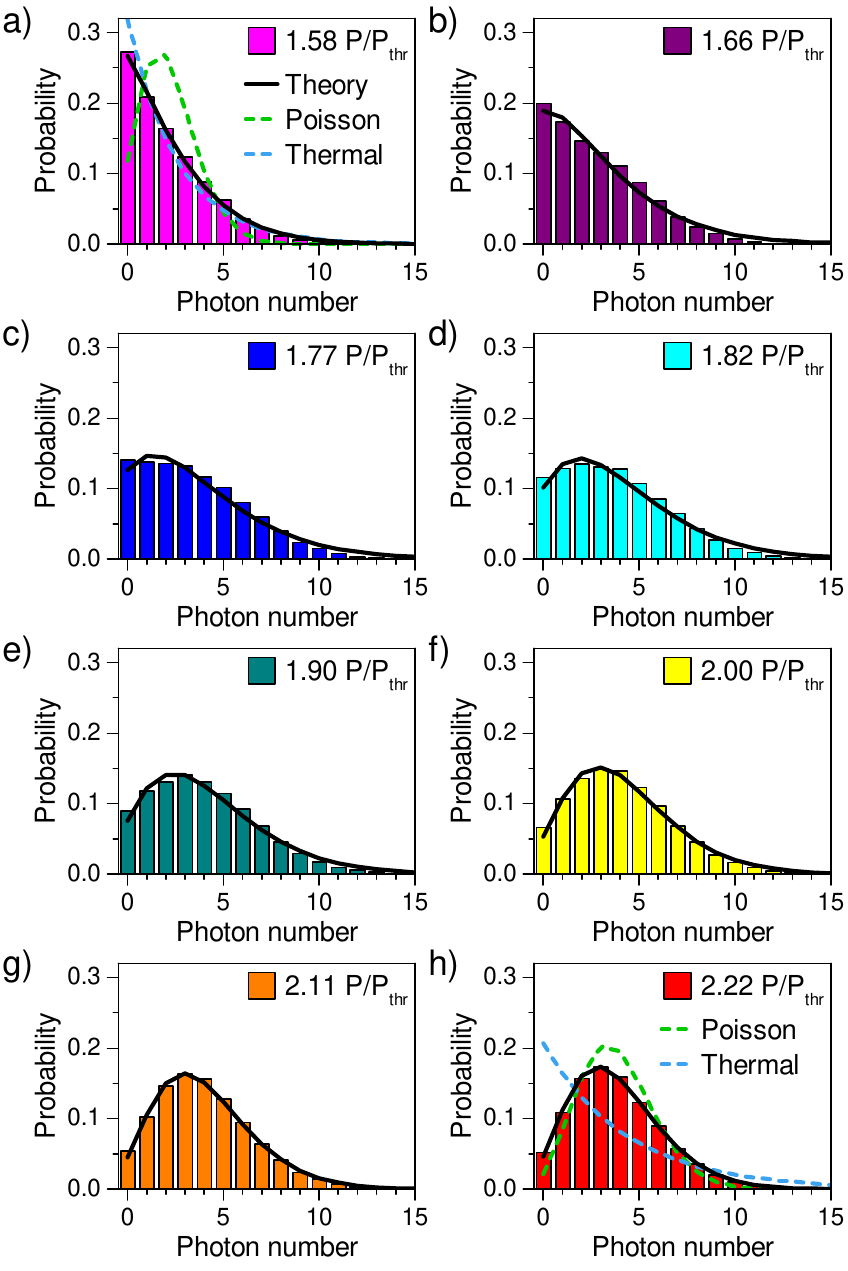}
\caption{a)-h) Evolution of the photon-number distribution with increasing excitation power. a) Shows a near-exponential dependency, signifying thermal emission, while b) to h) exhibit a transition between dominating thermal to mainly coherent proportions of a mixed state. h) Depicts a quasi Poissonian distribution; a laser like emission state. All measurements have been fitted with a thermal-coherent transition state shown by the black lines. a) and h) additionally show dashed line plots of pure thermal (blue) and pure Poissonian statistics (green) of the same $\langle \hat{n} \rangle$ as the experimental data.}
\label{Fig2}
\end{figure}

Figure \ref{Fig2} shows the experimental photon-number distributions together with a theory fit (details are given in theory section) for different excitation powers (1.58 - 2.22 times $P_{th}$). At a moderate pump power, relative to the onset of the intensity nonlinearity, the emission has an exponential-like photon-number distribution and resembles a thermal emitter [see panel \ref{Fig2} a)]. The panels a) to h) of Fig. \ref{Fig2} correspond to increasing excitation power. It is clearly seen that this system features a transition between the two emission regimes resulting in a combination of thermal and coherent emission. Lastly, in panel h) the system has reached a nearly coherent laser-like state with a quasi-Poissonian photon-number distribution. We now turn to a more detailed analysis of this experimental data.

\begin{figure}[tbp]
\centering
\includegraphics[width=\linewidth]{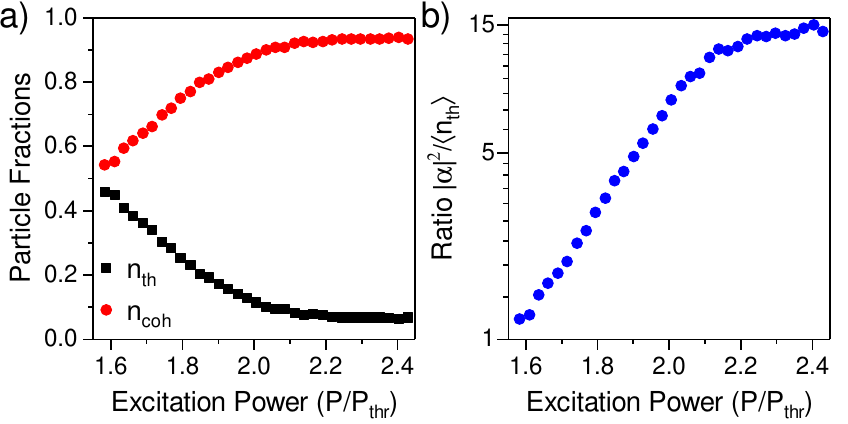}
\caption{a) Evolution of the thermal and coherence average particle fractions extracted from the fitted thermal-coherent state distributions presented in Fig. \ref{Fig2}. b) Ratio of the coherent and thermal populations across the condensation threshold.}
\label{Fig3}
\end{figure}

These photon-number distributions can be more quantitatively analyzed in terms of coherent thermal states \cite{OzVogt1991}. Such states are obtained by the application of a displacement operator $\hat D\left( \alpha  \right) = \exp \left( {\alpha {{\hat a}^\dag } - {\alpha ^*}\hat a} \right)$ of a complex parameter $\alpha$ to a thermal state characterized by the density matrix ${{\hat \rho }_{\rm th}} = \left( {1 - p} \right)\sum\limits_n {{p^n}\left| n \right\rangle } \left\langle n \right|$ written in a Fock state ${\left| n \right\rangle }$ basis and where $p = {\langle n_{th} \rangle}/\left( {{\langle n_{th} \rangle} + 1} \right)$ given $\langle n_{th} \rangle$ the mean thermal occupation. The coherent thermal state density matrix is therefore obtained as $\hat \rho(\alpha,\langle n_{th} \rangle)  = \hat D\left( \alpha  \right){{\hat \rho }_{\rm th}}{{\hat D}^\dag }\left( \alpha  \right)$ and its probability distribution is given by $P_n={\rm diag}[\hat \rho(\alpha,\langle n_{th} \rangle)]$. The thermal state $\alpha=0$ and the coherent state $\langle n_{th} \rangle=0$ probability distributions are respectively exponential and Poissonian in the photon-number $n$ with $P_n^{\rm th} = \frac{1}{{1 + {\langle n_{th} \rangle}}}{\left( {\frac{{{\langle n_{th} \rangle}}}{{1 + {\langle n_{th} \rangle}}}} \right)^n}$ and $P_n^{\rm co} = {e^{ - {{\left| \alpha  \right|}^2}}}\frac{{{{\left| \alpha  \right|}^{2n}}}}{{n!}}$. The general case reads \cite{Arecchi.1966}
\begin{eqnarray}
P_n=\frac{\langle n_{th} \rangle^n}{(1+\langle n_{th} \rangle)^{n+1}}e^{-\frac{\left| \alpha \right|^2 }{1+\langle n_{th} \rangle}}L_n\left(\frac{-\left| \alpha \right|^2}{\langle n_{th} \rangle+\langle n_{th} \rangle^2}\right).
\end{eqnarray}
with $L_n$ being the Laguerre polynomials of order $n$, its mean occupation is $\left\langle {\hat {n}} \right\rangle  = {\langle n_{th} \rangle} + {\left| \alpha  \right|^2}$ with $\langle {\Delta {{\hat n}^2}} \rangle  = {\left| \alpha  \right|^2}\left( {2{\langle n_{th} \rangle} + 1} \right) + \langle n_{th} \rangle^2 + {\langle n_{th} \rangle}$. We show in Fig.~\ref{Fig2} fits of our measured distributions to the ones of such coherent thermal states with black lines. We are then able to extract the corresponding coherence $|\alpha|^2$ and thermal ${\langle n_{th} \rangle}$ fractions (fit parameters) versus the pump power in Fig. \ref{Fig3}a). We observe a continuous drop of the average thermal fraction due to the condensation mechanism such that for high excitation powers, ${\langle n_{th} \rangle}$ nearly vanishes in favor of $\left|\alpha\right|^2$. Figure~\ref{Fig3}b) plots the ratio of $\left|\alpha\right|^2$ and $\langle n_{th} \rangle$. A rapid exponential increase is observed in the phase transition above threshold until a plateau, with a ratio of 15 is reached. It happens in presence of a small final persisting thermal occupation, reflecting the $g^{(k)}(0)$ behavior.

\begin{figure}[tbp]
\centering
\includegraphics[width=\linewidth]{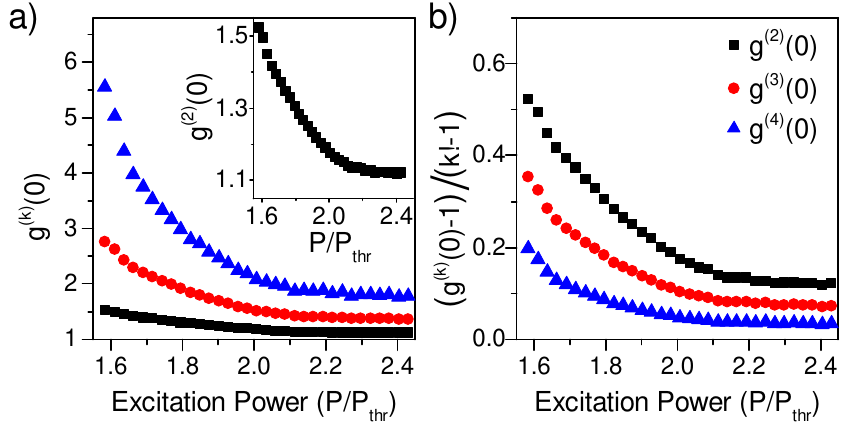}
\caption{a) Higher order photon correlations of a polariton condensate, evaluated from the photon-number distributions via the higher order momenta of the statistics. The inset shows a zoom in on the $g^{(2)}(0)$ measurement. b) $g^{(k)}(0)$ functions scaled with $(g^{(k)}(0)-1)/(k!-1)$.}
\label{Fig4}
\end{figure}

Beyond the basic evaluation of the photon-number distribution, the TES data also allows to directly reconstruct the photon correlations of the light source. Figure \ref{Fig4}a) exhibits the calculated autocorrelations for $g^{(2)}(0)$ to $g^{(4)}(0)$, extracted from distributions at each pump power. The higher-order autocorrelations up to the third order qualitatively confirm previous results, obtained from polariton devices with a significantly lower quality factor \cite{{Horikiri.2010, Assmann.2010}} and without any lateral confinement. The pronounced drop of $g^{(2)}(0)$ towards unity with increasing pump power above the threshold of polariton condensation indicates the buildup of a coherent state. Interestingly, with this measurement technique, we are able to determine even higher order autocorrelations, theoretically up to the order of the highest measured photon-number. 

Figure~\ref{Fig4}b) compares the correlation functions scaled as $\frac{g^k(0)-1}{k!-1}$ \cite{Assmann.2009}. It transforms the highest-order photon correlations in such a way that a thermal state corresponds to one and a coherent state to zero for each $g^{(k)}(0)$. This allows us to relate autocorrelation functions to each other. While the general decay shape from a thermal to a coherent state is preserved, the higher order photon correlations show a lower deviation from the coherent state for all excitation powers. Previously, Ref. \cite{Assmann.2009} reported the opposite for higher order photon correlations. However, the results from Ref. \cite{Horikiri.2010}, using the same scaling, happen to agree with our measurement. The sample of Ref. \cite{Assmann.2009} shows a transition from strong to weak coupling at an unknown input power. The $g^{(k)}(0)$ behavior is therefore expected to differ compared to persisting strong coupling conditions. 

The textbook coherent-thermal states theory stands as a very good fit for the experimental data in Fig. \ref{Fig2} but displays slight deviations for the lower excitation powers in the high photon-numbers. These small discrepancies, most markedly pronounced at the transition, should provide deeper insights into the exact coherence formation mechanisms and merit further future detailed theoretical study to attribute them to an either more laser-like or more atom-like phase transition in the careful investigation of different models like the weakly and strongly interacting Bose gas. At this stage of our investigation and the current sample, we find that the condensation follows closely the paradigm of a laser. This is consistent with the fact that our sample exhibits photon-like condensation. Interesting further studies involve the differing behaviors in other optical accessable phase transitions, like purely photonic condensates \cite{Klaers.2010}, polariton condensates in equilibrium in ultra-high quality samples \cite{Sun.2017a} and even Frenkel exciton condensates for which different interactions are proposed \cite{Dietrich.2016}. Samples with very strong polariton interactions \cite{Sun.2017b, Rosenberg.2018} will also allow fascinating explorations of strong-correlations through the full particle-number distribution. With the study and comparison of these systems, this powerful new photon-number resolving sensor enables to investigate in a new light the role of the interaction present at differing strength levels for each mentioned system. It further allows for even more sophisticated measurement schemes to access the photon statistics in different sample and emission configurations (e.g. to map the exact contributions of higher energy states by careful filtering to produce more complex convoluted multi-mode statistics). 

In conclusion, we have determined the photon statistics evolution at the phase transition of a polariton condensate in a strongly coupled microcavity via a transition edge sensor. Above threshold, the photon statistics changes from a thermal to a coherent distribution. This behavior can additionally be monitored with high-order photon correlation functions at zero delay which can be straightforwardly extracted from the photon statistics. This new measurement of the full statistical information for polaritons gives unique insights into the nature of their phase transition that should stimulate the development of competing theories to identify the mechanism at play in their coherence build-up.

\begin{acknowledgments}
The authors would like to thank the ImPACT Program, Japan Science and Technology Agency and the State of Bavaria for financial support. We thank Monika Emmerling and Adriana Wolf for expert sample processing. We thank NIST and PTB for providing the detector. C.S. thanks DFG project Schn1376/3-1. S. R. acknowledges funding from the European Research Council under the European Union's Seventh Framework ERC Grant Agreement No. 615613, the EURAMET joint research project MIQC2 from the European Union’s Horizon 2020 Research and Innovation Programme and the EMPIR Participating States, and the German Research Foundation within the project RE2974/10-1. F.P.L. acknowledges support from the Ministry of Science and Education of the Russian Federation (RFMEFI61617X0085).
\end{acknowledgments}

\end{document}